\documentclass[12pt]{article}

\usepackage[english]{babel}
\usepackage[utf8x]{inputenc}
\usepackage[T1]{fontenc}
\usepackage[prependcaption,textsize=tiny]{todonotes} 
\usepackage{bm} 
\usepackage[tight-spacing=true]{siunitx} 
\usepackage{tabularx} 
\usepackage{booktabs}
\presetkeys{todonotes}{inline,backgroundcolor=yellow}{}
\usepackage[colorlinks=true, allcolors=blue]{hyperref}

\title{Blue-blocking spectacles lenses for   retinal damage protection and circadian rhythm: evaluation parameters}

\author{Regina Comparetto$^1$ \and Alessandro Farini$^2$}
\date{%
{\small
$^1$University of Florence, degree in Optics and Optometry\\%
$^2$CNR-National Institute of Optics, Largo Enrico Fermi 6, 50125 Firenze\\[4ex]%
}
\today
}

\begin{document}
 \maketitle

\begin{abstract}
There is evidence for the effect of blue light on circadian cycle and ocular pathologies.  Moreover the introduction of LED lamps has increased the presence of blue light. In the last two years many different blue blocking ophthalmic  lens have been introduced Due to the different effect of blue light on ocular media and circadian rhythm we have defined two indices that   describe the level of protection of the lenses towards the retina and the circadian cycle under different lighting as for example daylight and tablet. These indexes can help in individuating the right lens for ocular protection.
\end{abstract}


\section{Introduction}
The invention of the Light Emitting Diode (LED) has completely changed the world of light sources.
LEDs have replaced the old type of light sources, and they have been massively adopted in the production of screens of all kind of electronic devices with which many of us deal everyday, such as TVs, PCs, smartphones, and tablets.
A particularity of the commonly used white LED is its significant emission of short-wavelength blue light \cite{nakamura2000introduction}. Thus, it is important to fully understand the consequences of this type of radiation on our health.

Recent investigations on the \textit{third retinal receptor} have shed some light on the importance of \textit{blue light} (i.e. short-wavelength visible radiation mainly in the range
400 nm $\leq \lambda \leq$ 450 nm) for our life cycles.
Berson et al. \cite{berson2002} showed that this receptor has a huge impact on the control of the circadian cycle, i.e. the set of all the physiological cycles of our body within 24 hours, such as the regulation of arterial pressure or the production of melatonin.
Several studies confirmed that an unnatural exposition to short-wavelength radiation can have a negative impact on our health.
For example, an insufficient exposure to blue light during the day has been related to sleep disorders \cite{van2004lighting,ayaki2013improvements}, while a strong exposure to blue light during the evening is known to cause the inhibition of melatonin production which damages the quality of sleep \cite{wood2012,chang2015evening,van2015blue}.
Lots of studies \cite{noell1966retinal,ham1980nature,ham1978histologic,margrain2004blue, ham1979sensitivity,sparrow2001blue,sparrow2000lipofuscin,fletcher2008sunlight,mainster2006violet} have shown how an excessive and prolonged exposure to this kind of radiation can lead to the onset of ocular pathologies, such as senile macular degeneration or cataract.

Optical devices such as ophthalmic, contact, or intra-ocular lenses can be used as a protection from the effects of the blue radiation.
Due to the ubiquitous presence of LED devices, we witnessed an increasing interest around those aspects of blue light, with the result of the market introduction over the last few years of many blue-blocking ophthalmic lenses by lens manufacturers.
However, there is still a debate in the scientific community and no strict regulations on how short-wavelength visible light should be treated. 
Thus, it is not clear how lenses should protect us from the negative effects of the blue light (by blocking it) while preserving the positive effects on the circadian cycle.

In this work, we propose a novel approach to characterize the interaction of lenses with blue light through the introduction of two numerical indexes. Those are meant to quantify the behavior of lenses when exposed to the blue radiation and to estimate both the risks of retinal damage and disruption of the circadian cycle.
We evaluate a set of commercially available lenses with different types of blue-blocking treatments, comparing them with a group of lenses without this type of treatment.
Computing our indexes for the analyzed lenses, we were able not only to compare treated and non-treated lenses, but also to capture the heterogeneity of the behaviors of the different blue-blocking lenses.

The paper is structured as follows.
In section 2, we briefly introduce the numerical parameters used for the analysis of the lenses, giving the definition of the two novel indexes we propose in this work.
In section 3 we report the experimental settings, in section 4 we expose and discuss the results of this study, and section 5 concludes the paper.

\section{Characterization of the interaction of lenses with blue-light radiation}
In this section, we describe the two proposed indexes to characterize the response of lenses to the exposure to blue light, respectively named \textit{Retinal Index} (RI) and \textit{Circadian Index} (CI).

\paragraph{Retinal Index (RI)}

\begin{figure}
\centering
\includegraphics[width=\linewidth]{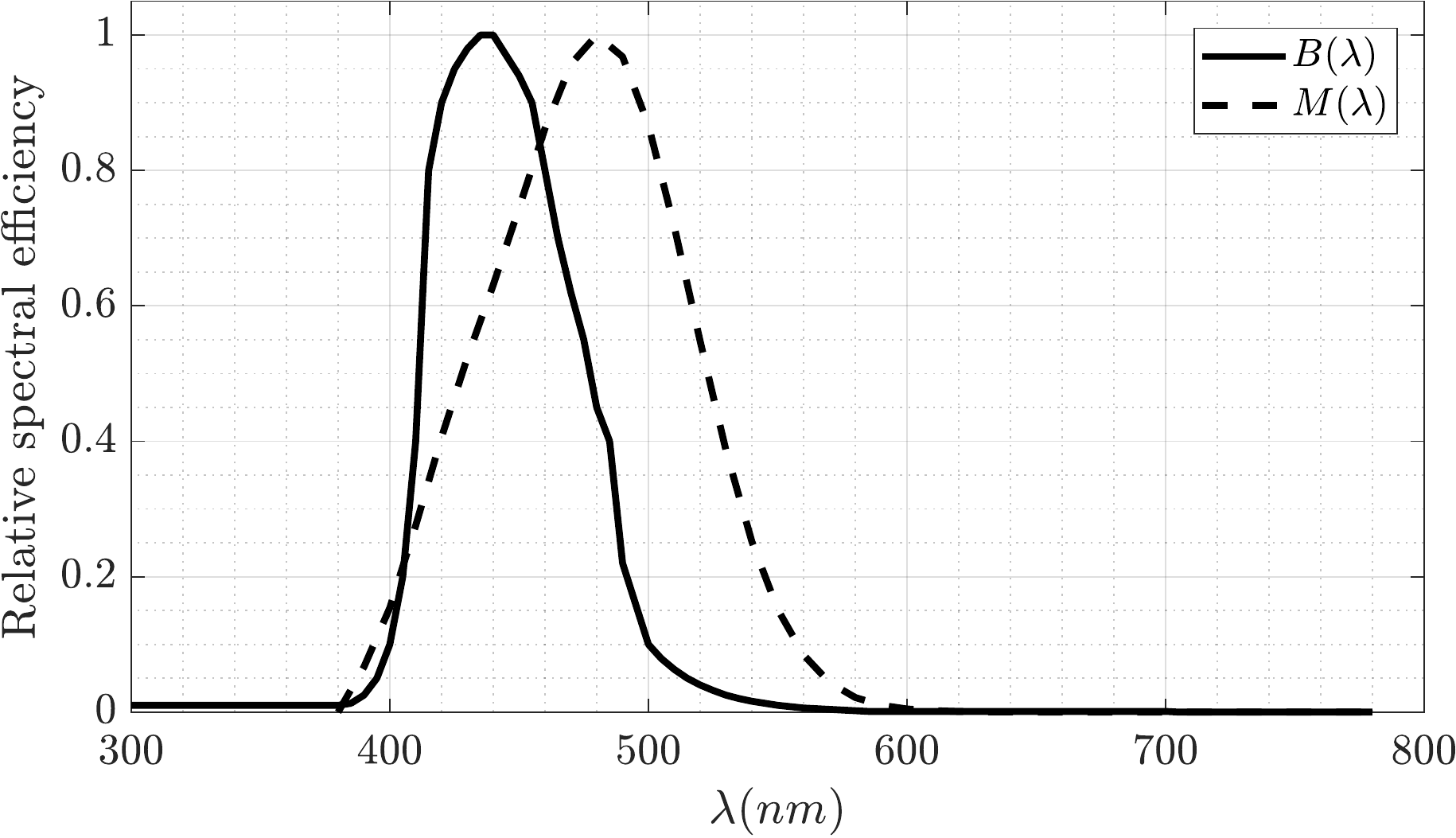}
\caption{Comparison between $B(\lambda)$, Blue Light Hazard Function, and $M_{\lambda}$, a relative spectral efficiency function which represents the represents the response of the third retinal receptor to radiation.}
\label{fig:MB}
\end{figure}

The Retinal Index (RI) quantifies the possible damage of the retina due to the exposure to the short-wavelength radiation. 
We define RI as:
\begin{equation} \label{eq:RI}
RI=\frac{\displaystyle \int_{380 nm}^{780 nm}{T(\lambda) I(\lambda) B(\lambda) d\lambda}}			{\displaystyle \int_{380 nm}^{780 nm}{I(\lambda)B(\lambda)d\lambda}}
\end{equation}
where $T(\lambda)$ is \textit{spectral transmittance} of a lens in the visible spectrum (380 -- 780 nm), defined as the ratio between the transmitted flux and the incident flux \cite{palmer1995measurement}, $I(\lambda)$ is a generic illuminant%
, and $B(\lambda)$ is the \textit{blue light hazard function}  (depicted in Fig. \ref{fig:MB}). $B(\lambda)$ represents the risk of damaging the retina if exposed to a blue-light radiation \cite{icnirp_guide}.
RI ranges from 0 to 1, where a RI = 0 identifies a totally protective lens against the photochemical retinal damage due to blue light, while RI = 1 identifies a totally non-protective lens.
	
\paragraph{Circadian Index (CI)}
	
The Circadian Index (CI) quantifies the ability of a lens to inhibit the effect of the light radiation in the circadian cycle.
We define CI as the weighted average of the spectral transmittance of the lens:
\begin{equation} \label{eq:CI}
CI=\frac{\displaystyle \int_{380 nm}^{780 nm}{T(\lambda)I(\lambda)M_{\lambda}d\lambda}}
    {\displaystyle \int_{380 nm}^{780 nm}{I(\lambda)M_{\lambda}d\lambda}}
\end{equation} 
where $T(\lambda)$ is the spectral transmittance, $I(\lambda)$ is the emission spectrum a generic illuminant,
and $M_{\lambda}$ (depicted in Fig. \ref{fig:MB}) is a relative spectral efficiency function that represents the response of the third retinal receptor to the light radiation \cite{dacey2005, rea2010}. The function $M_{\lambda}$ takes into account how much the circadian cycle is influenced by the received radiation.
CI ranges from 0 to 1, where a CI = 0 identifies a lens that completely blocks the effects of the blue light radiation on the circadian cycle, and  CI = 1 identifies a lens that does not interfere with those effects.
When using a lens with CI = 1, the natural circadian cycle could be altered by the exposure to the artificial blue light of digital devices, but the same lens allows the natural rhythm of the circadian cycle in the case of natural, solar light exposure.
On the other hand, a lens with CI = 0 protects from the damages caused by the artificial blue light, but, at the same time, does not allow the natural blue light to reach the third retinal receptor.

\paragraph{UV Transmission Factor ($\tau_{UV}$)}
To fully characterize the quality of a lens, we also measure and report the \textit{Solar UV transmission factor, $\tau_{UV}$}, defined as

\begin{equation} \label{eq:tauUV}
\tau_{UV}=\frac{\displaystyle \int_{280 nm}^{380 nm}{T(\lambda)W_{\lambda}(\lambda)d\lambda}}			{\displaystyle \int_{280 nm}^{380 nm}{W_{\lambda}(\lambda)d\lambda}}
\end{equation}

where $T(\lambda)$ is the spectral transmittance of a lens, and $W_{\lambda}(\lambda)$ is the weighting function for UV transmission as defined in (European Regulation UNI EN 1836) \cite{}.
This index takes into consideration the percentage of ultraviolet radiation that a medium is able to transmit.

\section{Experimental Settings}

\begin{table}

\caption{List of all the analyzed lenses. When available, we also report their refractive indexes ($\bm{n}$), whether they had a blue blocking treatment (\textbf{BB}), and the RI and CI computed with the standard illuminant D65 ($\bm{RI_{D65}}$, $\bm{CI_{D65}}$) and the spectral emission of a LCD screen ($\bm{RI_{LCD}}$, $\bm{CI_{LCD}}$).}

\label{tab:lenti}

\newcolumntype{e}{%
    S[{
        table-number-alignment=center,
        round-mode=places,
        round-precision=2
    }]%
}
\newcolumntype{u}{%
    S[{
        table-number-alignment=center,
        round-mode=places,
        round-precision=1
    }]%
}
\newcolumntype{C}{>{\centering}X}

\begin{tabularx}{\linewidth}{CCCeeeeu}
\bottomrule
\textbf{Lens}        & $\bm{n}$ & \textbf{BB} & $\bm{RI_{D65}}$ & $\bm{CI_{D65}}$ & $\bm{RI_{LCD}}$ & $\bm{CI_{LCD}}$ & $\bm{\tau_{UV}\%}$ \\
\midrule
1                    & 1.50           & y         & 0.767355367491019   & 0.837470660741561   & 0.778788750530242    & 0.838136208914397    & 3.099023204806948 \\
2                    & 1.50           & y         & 0.857923002922294   & 0.917216609491585   & 0.918434214524021    & 0.934704418101468    & 3.286431234623702 \\
3                    & -              & y         & 0.864857985067335   & 0.918472535055787   & 0.882764377811450    & 0.922287584887496    & 1.737975373676655 \\
4                    & -              & y         & 0.900637923082859   & 0.929706900703799   & 0.901463921140052    & 0.929529628749974    & 4.863678039180964 \\
5                    & -              & y*        & 0.099550708044133   & 0.229908988430876   & 0.093876978017768    & 0.241043190012744    &  0.007071162305891\\
6                    & 1.50           & n          & 0.844019995408132   & 0.854698229526632   & 0.847591376048842    & 0.853863854570201    & 4.762423404437660 \\
7                    & 1.59           & y         & 0.793700596710798   & 0.861837094507608   & 0.810756329010041    & 0.865270371824413    & 0.034202171740000 \\
8                    & 1.74           & y         & 0.790896735647903   & 0.866709450219304   & 0.850974790016966    & 0.893546398167232    & 0.006029784213449 \\
9                    & 1.60           & y         & 0.739665254999050   & 0.818661232054486   & 0.774641067910206    & 0.834481353461866    & 0.006053793777643 \\
10                   & 1.67           & y         & 0.783558420085130   & 0.862174867379096   & 0.833558041396773    & 0.883259356700238    & 0.012520847675742 \\
11                   & 1.59           & y         & 0.850435803287146   & 0.905827322871341   & 0.918143375212265    & 0.932343911733474    & 0.003026469981322 \\
12                   & 1.60           & y         & 0.825579822298609   & 0.892751861978832   & 0.909754057396000    & 0.928045417219131    & 0.001919908916873 \\
13                   & 1.67           & y         & 0.818227878541686   & 0.886353009585475   & 0.906299663137232    & 0.923352854282491    & 0.002679310515954 \\
14                   & 1.50           & y         & 0.838519854959740   & 0.898798128183588   & 0.896530046666761    & 0.926286521680683    &  0.002258314352504\\
15                   & 1.50           & y         & 0.754393564032212   & 0.866943023897008   & 0.918519741815361    & 0.930649036955708    &  0.015416747657491\\
16                   & 1.60           & y         & 0.802025943310001   & 0.896114437394572   & 0.958366562208832    & 0.962443100424260    &  0.001667823969121\\
17                   & 1.60           & y         & 0.819809402312451   & 0.906535643916307   & 0.965753638103811    & 0.969702602586979    &  0.002817143144496\\
18                   & -              & n          & 0.965226388220591   & 0.971167507454409   & 0.975781387606741    & 0.976900447387971    & 5.221003734742137 \\
19                   & 1.50           & n          & 0.884183779011212   & 0.900402496717376   & 0.892642592475971    & 0.903351016602616    &  2.845686371392069\\
20                   & 1.50           & n          & 0.869999273554703   & 0.886602128089070   & 0.877277820689021    & 0.888911604150331    & 2.811624290560454 \\
21                   & 1.50           & n          & 0.920302009616224   & 0.920358131151534   & 0.920232722084437    & 0.920310110317024    &  66.681575883590950$^\dagger$\\

\midrule
\end{tabularx}

* Lens 5 is an orange-tinted lens. \\
$^\dagger$ Lens 21 is a non-organic glass lens without any treatment.
\end{table}

For this study, we analyzed 16 commercially available blue-blocking lenses from 8 different companies and 5 lenses without any blue-blocking treatment. The characteristics of the analyzed lenses are reported in Table \ref{tab:lenti}.

We measured the spectral transmittance $T(\lambda)$ using the spectrophotometer \textit{Perkin Elmer Lambda 1050 UV/Vis/NIR} double-bean with integrating sphere.
Since we were interested in the behavior of the lenses in the visible and UV spectra, we measured $T(\lambda)$ for 280 nm $\leq \lambda \leq$ 830 nm, with a stride of 5 nm, in order to obtain a good compromise between measuring times and preciseness, as suggested by the main ISO regulations.
For each lens, we computed RI, CI, and $\tau_{UV}$ using two different illuminants $I(\lambda)$, the standard illuminant D65 ($RI_{D65}$ e $CI_{D65}$), and the spectral emission of a LCD screen, in particular the one of an iPad ($RI_{LCD}$ e $CI_{LCD}$).
By changing the illuminant it's possible to study the behavior of a medium when exposed to different type of radiation.
The MATLAB code for the computation of the indexes is publicly available\footnote{\url{https://github.com/ReginaComparetto/Retinal-and-Circadian-Indexes}}.

\section{Results and discussion}

\begin{figure}
\centering
\includegraphics[width=\linewidth]{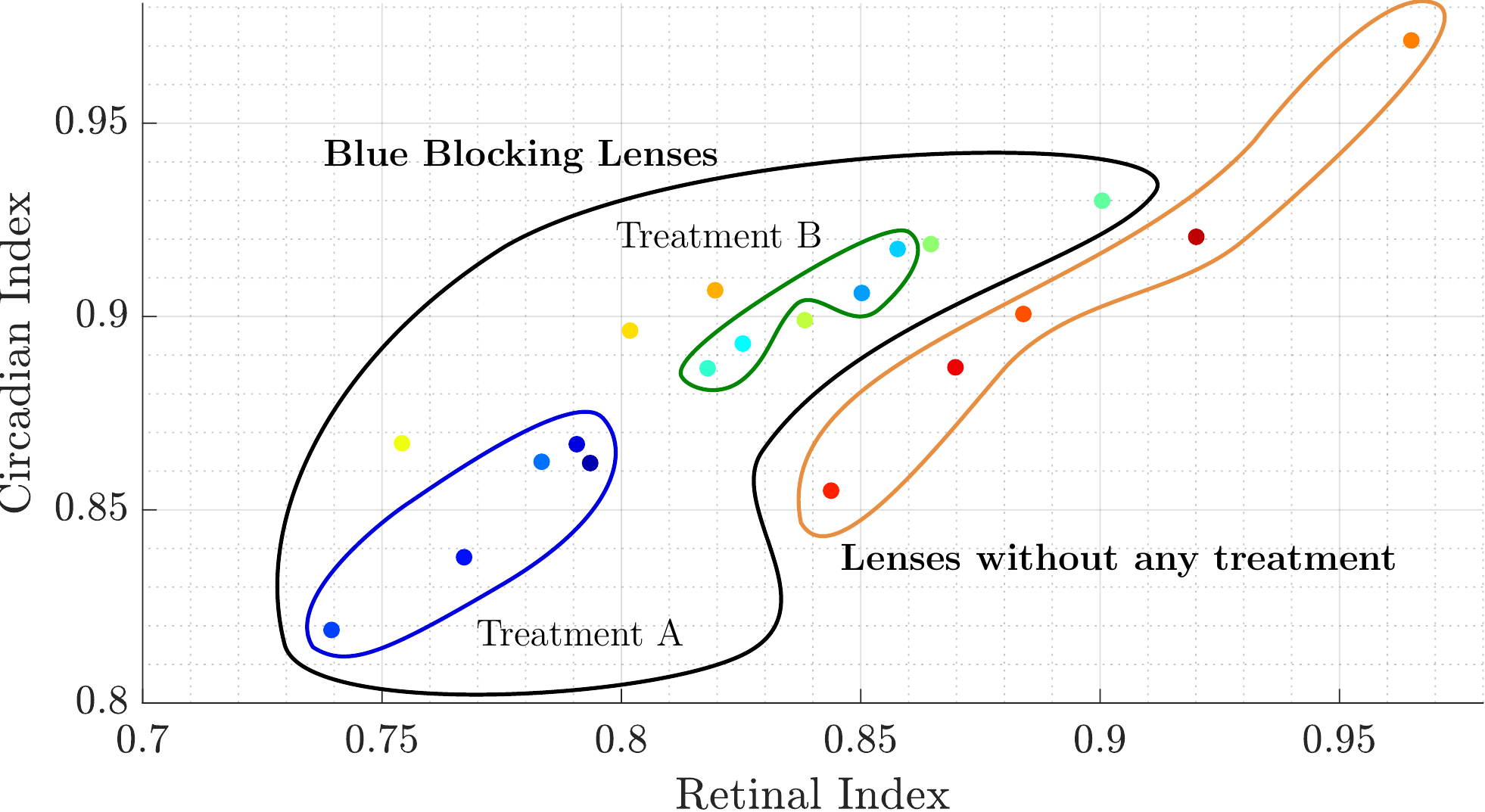}
\caption{Scatter plot of RI and CI some of the evaluated lenses. The reported indexes are calculated using the standard illuminant D65.}
\label{fig:RICI}
\end{figure}

The computed indexes for all the lenses and both the illuminants are reported in Table \ref{tab:lenti}.
Figure \ref{fig:RICI} shows the scatter plot of RI and CI of various lenses computed using the spectral emission of the standard illuminant D65.

Although it is clear that blue-blocking lenses are on average more protective towards the effects of blue light with respect to non-treated ones, we notice a heterogeneity in the RI and CI values among treated lenses that is not reported by lens manufactures.
For example, lenses with a different CI should be used for different needs, e.g. lower CI lenses might be used with electronic devices in the evening to prevent sleep disorders, while they are not recommended in the day-light in order to preserve the natural circadian cycle. Instead, lens manufactures generally advertise lenses with blue-blocking treatments as globally protective against blue light without distinguish those aspects. 

\paragraph{Indexes Correlation}
We can observe that there is a positive correlation between the two indexes. This is reasonable since the peak values of $B(\lambda)$ and $M_\lambda$ (on which the definition of RI and CI are based) are near in the spectrum, and there is a significant overlap of the areas under the two curves (see Fig. \ref{fig:MB}).
This means that the two aspects of the blue light, i.e. the effect on the circadian cycle and the damages of the retina, are difficult to separate.

\begin{figure}[htbp]
\centering
\includegraphics[width=\linewidth]{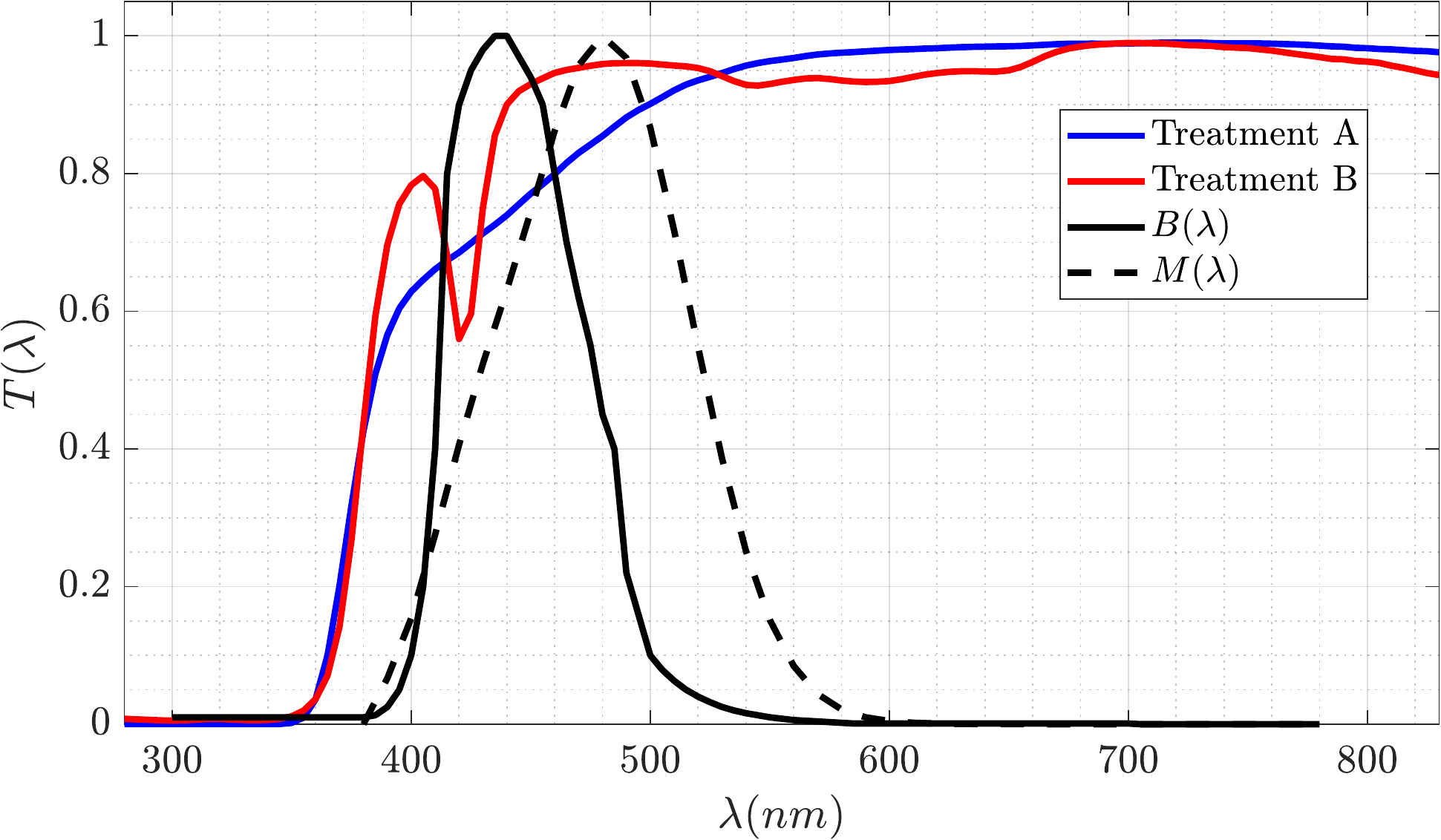}
\caption{Comparison between $B(\lambda)$, $M_{\lambda}$, and the spectral transmittance $T(\lambda)$ of two blue-blocking lenses (Lens 1 and 2) of the same material with different treatment (Treatment A and Treatment B highlighted in Figure \ref{fig:RICI}).}
\label{fig:CPEP_MB}
\end{figure}

In Figure \ref{fig:CPEP_MB}, the two curves $B(\lambda)$ and $M_{\lambda}$ are shown together with the spectral transmittance of two chosen lenses (Lens 1 and 2).
These lenses are two samples of the same material and made by the same company, but with two different blue-blocking treatments, which we refer to as \textit{Treatment A} and \textit{Treatment B}.
We can observe that the spectral transmittance of the lens with Treatment B reaches a high value near the higher point of $M_{\lambda}$ curve (450 nm $\leq\lambda\leq 500 nm$); thus, it has a higher $CI_{D65}$ than the lens with Treatment A.
On the other hand, a side effect of Treatment B is that the spectral transmittance near the peak of $B(\lambda)$ curve (400 nm $\leq\lambda\leq 450 nm$) is equally high, yielding a higher $RI_{D65}$ with respect to Treatment A.
However, we can notice that there is a clear depression in the spectral transmittance under the $B(\lambda)$ curve right before its peak, which prevents the retinal index to further increase.


\paragraph{Relation to UV transmittance}

\begin{figure}
\centering
\includegraphics[width=.95\linewidth]{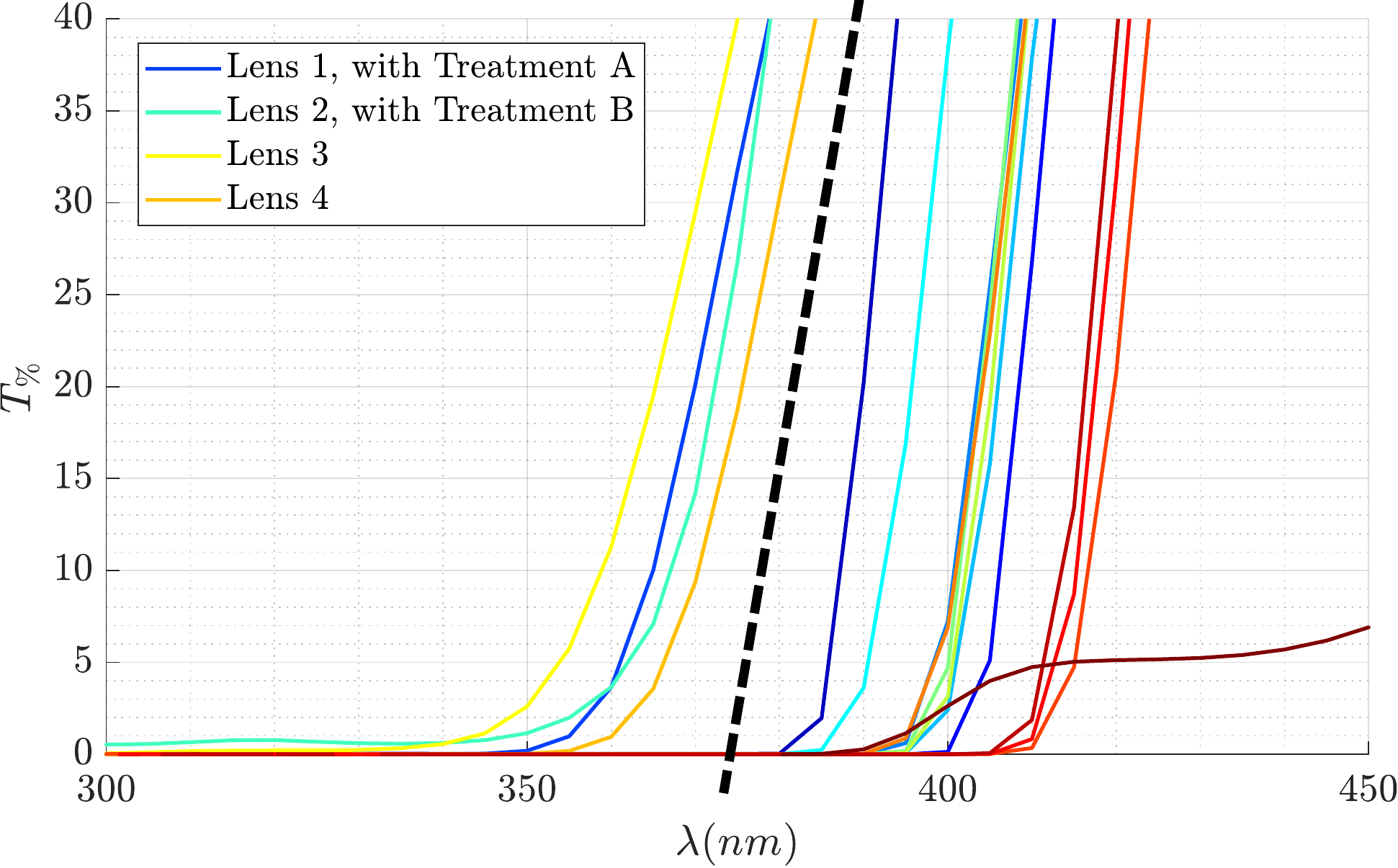}
\caption{The detail of the spectral transmittance of all blue-blocking lenses around the cut-off wavelength $\lambda_{cut}$.}
\label{fig:lambdataglio}
\end{figure}

In Figure \ref{fig:lambdataglio}, the spectral transmittances of all the analyzed lenses with blue blocking treatment are shown, with particular attention to the cut-off wavelength. 
We notice that four lenses (Lens 1 to 4) have a lower cut-off wavelength and thus present a higher UV transmission factor $\tau_{UV}$.
Despite being advertised as a protective lens with blue-blocking treatment, Lens 4 have the highest values of $RI_{D65}$, $CI_{D65}$, and $\tau_{UV}$ among all lenses with blue-blocking treatment; thus it offers a very limited protection against the effects of the exposure to blue light.
Moreover, it is the less protective less against UV radiation among all the treated ones, with a $\tau_{UV}$ value comparable to the ones of non-treated lenses.
Lens 1, which is the one with the Treatment A, has a lower value of $RI_{D65}$ and $CI_{D65}$, but still has a high value of $\tau_{UV}$, which is comparable to the one of lenses without blue blocking treatment.
It is important to notice that the values of RI and CI are independent from $\tau_{UV}$; thus the proposed indexes are not meant to describe in any way the behavior of the sample to the ultraviolet radiation.


\paragraph{Chromaticity of Lenses}

\begin{figure}
\centering
\includegraphics[trim={1.1cm 0.3cm 1.3cm 1.7cm},width=.9\linewidth]{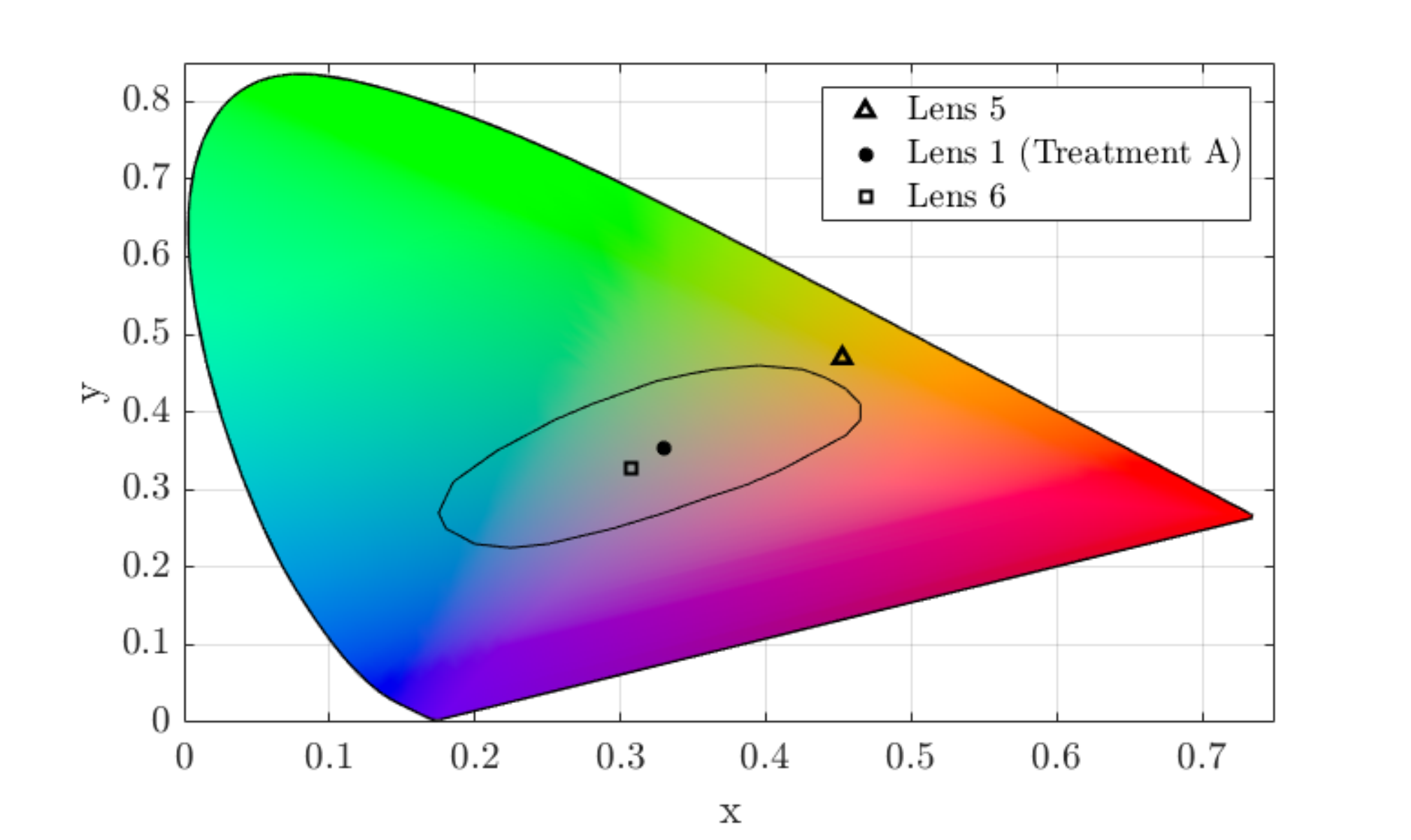}
\caption{The CIE color space chromaticity diagram. The points represented three out of all the analyzed lenses (with and without blue blocking treatment. The CIE spectral locus was generated using the software described in \cite{westland2012computational}}
\label{fig:CIE}
\end{figure}

In Figure \ref{fig:CIE}, we highlight the chromatic coordinates of three lenses of interest in the CIE color space chromaticity diagram.
The curve in the center of the diagram delimits the set of chromatic coordinates for which the color perception is not altered \cite{xylim_normativa1977}.
We observed that lenses with blue-blocking treatments do not lead to an altered color perception; in fact, all blue-blocking lenses fall inside the delimited area.
Lens 1, treated with Treatment A, is the most yellow lens out of all the one in the acceptance area; still, it is very near in the chromaticity space to Lens 6, a non-treated lens which is the most white of all the analyzed lenses.
The only lens that lies outside the acceptance area is Lens 5. Since it is an orange-tinted lens, we expected it to alter the perception of color. This accounts also for the low values of $RI_{D65}$ and $CI_{D65}$, since the strong orange tint blocks the majority of the blue radiation.
All the other lenses had color coordinates that lie approximately between the coordinates of Lens 1 and Lens 6.

\section{Conclusions}

In this work, we proposed two numerical indexes, namely the Circadian Index and the Retinal Index, to quantify the effects of the exposure to short-wavelength visible radiation to the human health.
Given a lens, the former summarizes impact of the transmitted light on the circadian cycle, while the latter summarizes the risk of retinal damage when exposed to same transmitted radiation.


Using these indexes, we performed a comparative analysis between commercially available lenses with blue-blocking treatment and non-treated lenses.

Results shown there is a large dispersion of behaviors among different treatments. Our proposed indexes are able to efficiently capture those differences, and they could be useful as a metric to characterize blue-blocking optical media.

Differently from already proposed index measuring global blue light transmission, we argue that having two separated metrics could help to easily identify the optimal lens for a particular usage. While it is always desirable to have a lens protecting from retinal damage, i.e. with a low RI, we may want to choose whether to block the effects of blue light to the circadian circle (with a low CI) or not to (with a high CI), depending on the needs of the user.

In order to encourage further research in this field, we released the data and MATLAB code to compute the proposed indexes and to replicate the experimental results.

\bibliographystyle{unsrt}
\bibliography{sample}



\end{document}